\documentclass[twocolumn]{aastex631}
\usepackage{xspace}

\newcommand{\ha}{H$\alpha$\xspace}
\newcommand{\mstar}{M$_*$\xspace}
\newcommand{\msun}{M$_\odot$\xspace}
\newcommand{\twister}{Twister-z5}

\begin{document}

\title{FRESCO: An extended, massive, rapidly rotating galaxy at $z=5.3$}

\author[0000-0002-7524-374X]{Erica Nelson}
\affiliation{Department for Astrophysical and Planetary Science, University of Colorado, Boulder, CO 80309, USA}

\author[0000-0003-2680-005X]{Gabriel Brammer}
\affiliation{Cosmic Dawn Center (DAWN), Denmark}
\affiliation{Niels Bohr Institute, University of Copenhagen, Jagtvej 128, DK-2200 Copenhagen N, Denmark}

\author[0000-0001-9419-9505]{Clara~Gim\'{e}nez-Arteaga}
\affiliation{Cosmic Dawn Center (DAWN), Denmark}
\affiliation{Niels Bohr Institute, University of Copenhagen, Jagtvej 128, DK-2200 Copenhagen N, Denmark}

\author[0000-0001-5851-6649]{Pascal A. Oesch}
\affiliation{Department of Astronomy, University of Geneva, Chemin Pegasi 51, 1290 Versoix, Switzerland}
\affiliation{Cosmic Dawn Center (DAWN), Niels Bohr Institute, University of Copenhagen, Jagtvej 128, K\o benhavn N, DK-2200, Denmark}

\author[0000-0003-4891-0794]{Hannah \"Ubler}
\affiliation{Kavli Institute for Cosmology, University of Cambridge, Madingley Road, Cambridge, CB3 0HA, UK}
\affiliation{Cavendish Laboratory, University of Cambridge, 19 JJ Thomson Avenue, Cambridge, CB3 0HE, UK}

\author[0000-0002-2380-9801]{Anna de Graaff} \affiliation{Max-Planck-Institut f\"ur Astronomie, K\"onigstuhl 17, D-69117, Heidelberg, Germany}

\author[0000-0002-7547-3385]{Jasleen Matharu}
\affiliation{Cosmic Dawn Center (DAWN), Denmark}
\affiliation{Niels Bohr Institute, University of Copenhagen, Jagtvej 128, DK-2200 Copenhagen N, Denmark}

\author[0000-0003-3997-5705]{Rohan~P.~Naidu}
\altaffiliation{NASA Hubble Fellow}
\affiliation{MIT Kavli Institute for Astrophysics and Space Research, 77 Massachusetts Ave., Cambridge, MA 02139, USA}

\author[0000-0003-3509-4855]{Alice E. Shapley}\affiliation{Department of Physics \& Astronomy, University of California, Los Angeles, 430 Portola Plaza, Los Angeles, CA 90095, USA}

\author[0000-0001-7160-3632]{Katherine E. Whitaker}
\affil{Department of Astronomy, University of Massachusetts, Amherst, MA 01003, USA}
\affiliation{Cosmic Dawn Center (DAWN), Denmark}

\author[0000-0003-1657-7878]{Emily Wisnioski}
\affiliation{Research School of Astronomy and Astrophysics, Australian National University, Canberra, ACT 2611, Australia}
\affiliation{ARC Centre of Excellence for All Sky Astrophysics in 3 Dimensions (ASTRO 3D)}

\author{Natascha M. F\"orster Schreiber}
\affiliation{Max-Planck-Institut für extraterrestrische Physik, Giessenbachstrasse 1, D-85748 Garching, Germany}

\author[0000-0001-8034-7802]{Renske Smit}
\affiliation{Astrophysics Research Institute, Liverpool John Moores University, 146 Brownlow Hill, Liverpool L3 5RF, UK}

\author[0000-0002-8282-9888]{Pieter van Dokkum}
\affiliation{Astronomy Department, Yale University, 52 Hillhouse Ave,
New Haven, CT 06511, USA}

\author[0000-0002-0302-2577]{John Chisholm}
\affiliation{Department of Astronomy, The University of Texas at Austin, 2515 Speedway, Stop C1400, Austin, TX 78712-1205, USA}

\author[0000-0003-4564-2771]{Ryan Endsley}
\affiliation{Department of Astronomy, The University of Texas at Austin, 2515 Speedway, Stop C1400, Austin, TX 78712-1205, USA}

\author[0000-0002-5891-1603]{Abigail I. Hartley}
\affiliation{Department for Astrophysical and Planetary Science, University of Colorado, Boulder, CO 80309, USA}

\author{Justus Gibson}
\affiliation{Department for Astrophysical and Planetary Science, University of Colorado, Boulder, CO 80309, USA}

\author[0009-0004-3835-0089]{Emma Giovinazzo}
\affiliation{Department of Astronomy, University of Geneva, Chemin Pegasi 51, 1290 Versoix, Switzerland}

\author[0000-0002-8096-2837]{Garth Illingworth}
\affiliation{Department of Astronomy and Astrophysics, University of California, Santa Cruz, CA 95064, USA}

\author[0000-0002-2057-5376]{Ivo Labbe}
\affiliation{Centre for Astrophysics and Supercomputing, Swinburne University of Technology, Melbourne, VIC 3122, Australia}

\author[0000-0003-0695-4414]{Michael V. Maseda}
\affiliation{Department of Astronomy, University of Wisconsin-Madison, 475 N. Charter St., Madison, WI 53706, USA}

\author[0000-0003-2871-127X]{Jorryt Matthee}
\affiliation{Department of Physics, ETH Z{\"u}rich, Wolfgang-Pauli-Strasse 27, Z{\"u}rich, 8093, Switzerland}
\affiliation{Institute of Science and Technology Austria (IST Austria), Am Campus 1, Klosterneuburg, Austria}

\author[0000-0002-9672-3005]{Alba Covelo Paz}
\affiliation{Department of Astronomy, University of Geneva, Chemin Pegasi 51, 1290 Versoix, Switzerland}

\author[0000-0002-0108-4176]{Sedona H. Price}
\affiliation{Department of Physics and Astronomy and PITT PACC, University of Pittsburgh, Pittsburgh, PA 15260, USA}

\author[0000-0001-9687-4973]{Naveen A. Reddy}
\affiliation{Department of Physics and Astronomy, University of California, 
Riverside, 900 University Avenue, Riverside, CA 92521, USA}

\author[0000-0003-4702-7561]{Irene Shivaei}
\affiliation{Centro de Astrobiolog\'{i}a (CAB), CSIC-INTA, Ctra. de Ajalvir km 4, Torrej\'{o}n de Ardoz, E-28850, Madrid, Spain}

\author[0000-0001-8928-4465]{Andrea Weibel}
\affiliation{Department of Astronomy, University of Geneva, Chemin Pegasi 51, 1290 Versoix, Switzerland}

\author[0000-0003-3735-1931]{Stijn Wuyts}
\affiliation{Department of Physics, University of Bath, Claverton Down, Bath, BA2 7AY, UK}

\author[0000-0003-1207-5344]{Mengyuan Xiao}
\affiliation{Department of Astronomy, University of Geneva, Chemin Pegasi 51, 1290 Versoix, Switzerland}

\author[0000-0002-8909-8782]{Stacey Alberts} \affiliation{Steward Observatory, University of Arizona, 933 N. Cherry Avenue, Tucson AZ 85721 USA}

\author[0000-0003-0215-1104]{William M.\ Baker}
\affiliation{Kavli Institute for Cosmology, University of Cambridge, Madingley Road, Cambridge, CB3 0HA, UK}
\affiliation{Cavendish Laboratory, University of Cambridge, 19 JJ Thomson Avenue, Cambridge, CB3 0HE, UK}

\author[0000-0002-8651-9879]{Andrew J.\ Bunker} \affiliation{Department of Physics, University of Oxford, Denys Wilkinson Building, Keble Road, Oxford OX1 3RH, UK}
\author[0000-0002-0450-7306]{Alex J.\ Cameron} \affiliation{Department of Physics, University of Oxford, Denys Wilkinson Building, Keble Road, Oxford OX1 3RH, UK}

\author[0000-0003-3458-2275]{Stephane Charlot} \affiliation{Sorbonne Universit\'e, CNRS, UMR 7095, Institut d'Astrophysique de Paris, 98 bis bd Arago, 75014 Paris, France}

\author[0000-0002-2929-3121]{Daniel J. Eisenstein}
\affiliation{Center for Astrophysics $|$ Harvard \& Smithsonian, 60 Garden Street, Cambridge, MA 02138, USA}

\author[0000-0001-7673-2257]{Zhiyuan Ji} \affiliation{Steward Observatory, University of Arizona, 933 N. Cherry Avenue, Tucson AZ 85721 USA}

\author[0000-0002-9280-7594]{Benjamin~D.~Johnson}
\affiliation{Center for Astrophysics $|$ Harvard \& Smithsonian, 60 Garden Street, Cambridge, MA 02138, USA}

\author[0000-0002-0267-9024]{Gareth C.\ Jones} \affiliation{Department of Physics, University of Oxford, Denys Wilkinson Building, Keble Road, Oxford OX1 3RH, UK}

\author[0000-0002-4985-3819]{Roberto Maiolino} \affiliation{Kavli Institute for Cosmology, University of Cambridge, Madingley Road, Cambridge, CB3 0HA, UK}
\affiliation{Cavendish Laboratory, University of Cambridge, 19 JJ Thomson Avenue, Cambridge, CB3 0HE, UK}
\affiliation{Department of Physics and Astronomy, University College London, Gower Street, London WC1E 6BT, UK}

\author[0000-0002-4271-0364]{Brant Robertson} \affiliation{Department of Astronomy and Astrophysics University of California, Santa Cruz, 1156 High Street, Santa Cruz CA 96054 USA}

\author[0000-0001-9276-7062]{Lester Sandles}
\affiliation{Kavli Institute for Cosmology, University of Cambridge, Madingley Road, Cambridge, CB3 0HA, UK}
\affiliation{Cavendish Laboratory, University of Cambridge, 19 JJ Thomson Avenue, Cambridge, CB3 0HE, UK}

\author[0000-0002-1714-1905]{Katherine A. Suess}
\affiliation{Department of Astronomy and Astrophysics, University of California, Santa Cruz, 1156 High Street, Santa Cruz, CA 95064 USA}
\affiliation{Kavli Institute for Particle Astrophysics and Cosmology and Department of Physics, Stanford University, Stanford, CA 94305, USA}

\author[0000-0002-8224-4505]{Sandro Tacchella}
\affiliation{Kavli Institute for Cosmology, University of Cambridge, Madingley Road, Cambridge, CB3 0HA, UK}
\affiliation{Cavendish Laboratory, University of Cambridge, 19 JJ Thomson Avenue, Cambridge, CB3 0HE, UK}

\author[0000-0003-2919-7495]{Christina C.\ Williams} \affiliation{NSF’s National Optical-Infrared Astronomy Research Laboratory, 950 North Cherry Avenue, Tucson, AZ 85719 USA}

\author[0000-0002-7595-121X]{Joris Witstok} 
\affiliation{Kavli Institute for Cosmology, University of Cambridge, Madingley Road, Cambridge, CB3 0HA, UK}
\affiliation{Cavendish Laboratory, University of Cambridge, 19 JJ Thomson Avenue, Cambridge, CB3 0HE, UK}

\begin{abstract} 
With the remarkable sensitivity and resolution of JWST in the infrared, measuring rest-optical kinematics of galaxies at $z>5$ has become possible for the first time. This study pilots a new method for measuring galaxy dynamics for highly multiplexed, unbiased samples by combining FRESCO NIRCam grism spectroscopy and JADES medium-band imaging. Here we present one of the first JWST kinematic measurements for a galaxy at $z>5$. We find a significant velocity gradient, which, if interpreted as rotation yields $V_{rot} = 240\pm50$~km/s and we hence refer to this galaxy as Twister-z5. With a rest-frame optical effective radius of $r_e=2.25$kpc, the high rotation velocity in this galaxy is not due to a compact size as may be expected in the early universe but rather a high total mass, ${\rm log(M}_{dyn}/{\rm M}_\odot)=11.0\pm0.2$. This is a factor of roughly 4$\times$ higher than the stellar mass within $r_e$. We also observe that the radial H$\alpha$ equivalent width profile and the specific star formation rate map from resolved stellar population modeling is centrally depressed by a factor of $\sim1.5$ from the center to $r_e$. Combined with the morphology of the line-emitting gas in comparison to the continuum, this centrally suppressed star formation is consistent with a star-forming disk surrounding a bulge growing inside-out. While large, rapidly rotating disks are common to $z\sim2$, the existence of one after only 1Gyr of cosmic time, shown for the first time in ionized gas, adds to the growing evidence that some galaxies matured earlier than expected in the history of the universe.
\end{abstract}


\section{Introduction} \label{sec:intro}
Recently with the James Webb Space Telescope (JWST), very massive galaxy candidates have been discovered in the first billion years of cosmic history \citep{labbe:23,casey:23,akins:23}. This is surprising as $\Lambda$CDM cosmology predicts that galaxies form hierarchically from the merging of smaller galactic units. If confirmed spectroscopically, the existence of very massive galaxies so early is hard to reconcile with current models. In some cases, the stellar mass of these galaxies pushes up against the total number of baryons available in the most massive halos \citep[e.g.][]{boylan-kolchin:23}. Further buttressing this idea is the prodigious number of bright galaxies with photometric redshifts $z>10$ (or $<500$ Myr after the Big Bang), pointing again to a very early onset for galaxy evolution \citep[e.g.][]{Oesch:16,mason:22,naidu:22,finkelstein:22, castellano:22,bunker:23a,tacchella:23a}.

In addition to the surprising discovery at early times of massive and/or luminous galaxies is the discovery with JWST of apparently disk-dominated early galaxies \citep[e.g.][]{ferreira:22,nelson:23,robertson:23,baker:23,kartaltepe:23}. 
An abundance of disk-dominated galaxies at early cosmic time is unexpected in the context of studies based on projected axis ratios in $<1.6$\micron~imaging with HST which show that the fraction of galaxies with inferred disk-dominated morphologies drops dramatically at $z>2$ \citep[e.g.][]{vanderwel:14b,zhang:19}.
 It is also unexpected in the context of kinematic measurements showing that the majority of star-forming galaxies are less rotation-dominated at higher redshifts \citep[e.g.][]{forster-schreiber:06,forster-schreiber:09,forster-schreiber:18,wisnioski:11,wisnioski:15,wisnioski:19,gnerucci:11,kassin:12,miller:12,tacconi:13,simons:17,turner:17,johnson:18,ubler:19,price:20}.
These results are also surprising theoretically. First, in an expanding $\Lambda$CDM cosmology, the universe is expected to be much denser at early times than at later times. With more galaxies per unit volume, the rate at which galaxies interact with one another should theoretically be much higher.  In such a state of frequent bombardment, large disks would be unexpected. Second, owing to higher star formation efficiency or  stellar feedback efficiency in high star formation surface density  galaxies driving turbulence, early galaxies are expected to be more dispersion supported \citep[e.g.][]{ubler:19,pillepich:19,girard:21}. If more massive, more luminous, and larger disk galaxies really exist at early cosmic times, the universe may be able to form mature galaxies earlier than we thought. 

However, the implications of this early JWST work on our understanding of the early universe remain hazy owing to observational uncertainties. First, very massive galaxies at very early times may be neither massive nor particularly early because both their redshifts and stellar population properties are uncertain. Their photometry could be affected by active galactic nuclei (AGN), bursty star formation histories, and potentially yet unconsidered other oddities of the early universe \citep[e.g.][]{Endsley:22,Whitler:23,Kocevski:23,Mason:23,papovich:23}.
Second, the disky nature of these galaxies has also been inferred only from imaging. As this is just a two-dimensional projection of a three-dimensional object, there are degeneracies in the implied intrinsic shape. For instance, low axis ratios could reflect edge-on disks or prolate shapes \citep[e.g.][]{vanderwel:14b}. 

Determining how common massive disk galaxies are at early times requires spectroscopy, which allows us to measure redshifts, masses, and kinematics. However, owing to the wavelength coverage of previous facilities, measurements of ionized gas kinematics using rest-frame optical emission lines like \ha\ and [\ion{O}{3}] have thus far only been possible to $z<3.5$ \citep[e.g.][]{forster-schreiber:06,genzel:08,law:09,cresci:09,epinat:09,jones:10,wisnioski:11,wisnioski:12,wisnioski:15,Mancini:11,swinbank:12,stott:14,Leethochawalit:16,turner:17,price:20}. Studies using millimeter and radio telescopes have begun to trace kinematics at $z>3.5$ using e.g. the CII 158\micron\ or CO line which traces cooler gas with some finding significant rotation \citep[e.g.][]{hodge:12,Smit:18,Tadaki:19,neeleman:20,Tsukui:21,jones:21,pope:23,parlanti:23} and even surprisingly low velocity dispersions \citep[e.g.][]{rizzo:20,rizzo:21,rizzo:23,lelli:21,fraternali:21,xiao:22}.


With the launch of JWST, optical emission line kinematic measurements for galaxies at $z>4$ have become possible for the first time. 
Here we present \ha\ kinematics of a massive, rapidly rotating galaxy at $z=5.3$ using the NIRCam F444W grism. 
This galaxy is edge-on, extended, and is fortuitously aligned with the grism dispersion direction and hence provides a simpler test of this methodology than the typically smaller, fainter, less well-aligned galaxies which will require more sophisticated modeling \citep[as in e.g.][]{degraaff:23}. In \S2 we describe the observations and data reduction, in \S3 the methodology we use to measure kinematics and our kinematic results, and in \S4 the spatial distribution of the line emission relative to the continuum. Finally, in \S5 we discuss the implications of these results in the context of other recent work as well as the potential for using NIRCam grism spectroscopy to measure kinematics with JWST. In this paper, we assume the WMAP9 $\mathrm{\Lambda CDM}$ cosmology with $\Omega _{M}=0.2865$, $\Omega _{\Lambda}=0.7135$ and $H_0 = 69.32 \, \mathrm{km}~\mathrm{s}^{-1}~\mathrm{Mpc}^{-1}$ \citep{bennett:13}.
All magnitudes in this paper are expressed in the AB system \citep{oke:74}.

\begin{figure}[hbt]
    \centering
    \includegraphics[width=0.5\textwidth]{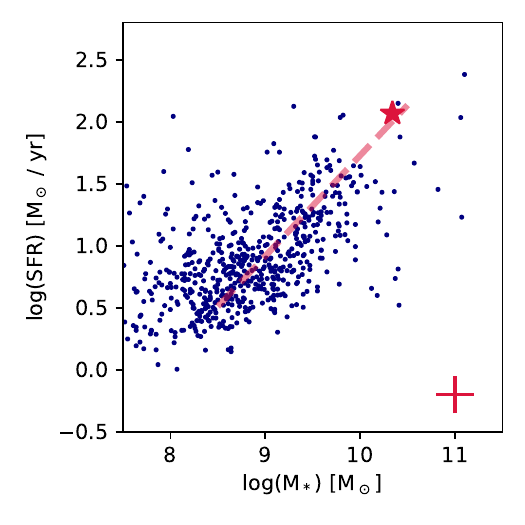}
    \caption{Star-forming main sequence for galaxies with grism spectroscopic redshifts $4.5<z<6$ in FRESCO with detected \ha. Star formation rates are computed from the measured \ha fluxes corrected for dust attenuation using empirical relations based on the UV slope \citep{shivaei:20}. A fit to the SFMS from \cite{speagle:14} is indicated by the red line for context. The typical measurement error is indicated by the cross in the lower right corner. \twister, the galaxy featured in the present study, is at the very massive end of this distribution on the upper end of a continuation of the locus of points.}
    \label{fig:sfms}
\end{figure}

\begin{figure*}
    \centering
    \includegraphics[width=\textwidth]{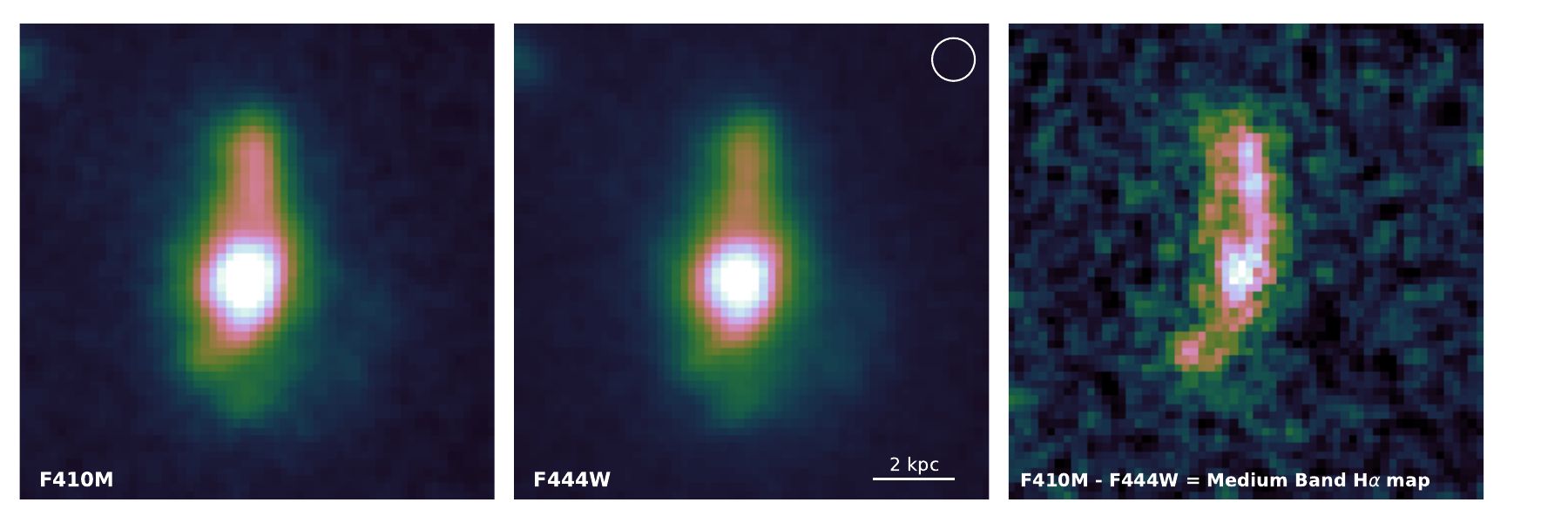}
    \caption{We infer the spatial distribution of \ha\ (right) from the difference between the F444W broadband filter (center) and F410M medium band filter (left) which covers the \ha\ emission line at this redshift. The circle in the middle panel shows the size (FWHM) of the PSF.}
    \label{fig:mbha}
\end{figure*}

\section{Data} \label{sec:data}
The data in this paper come from the FRESCO and JADES programs in GOODS-S \citep{oesch:23,rieke:23,eisenstein:23}. 
The object that forms the focus of our analysis which we refer to as \twister\ has coordinates [53.10169662, -27.83616465] and lies in the GOODS-S field. FRESCO is a 53.8 hour grism spectroscopic survey covering 124 arcmin${^2}$ in the GOODS-N and GOODS-S fields \citep{oesch:23}. It uses the NIRCam F444W grism, which has a maximal wavelength coverage of 3.8-4.9\micron\ and spectral resolution of R$\sim1600$ to measure  \ha\ [6563$\mathrm{\AA}$] at the end of the epoch of reionization. Observations in GOODS-S were taken between November 13 and 18, 2022, with an exposure time of 7ks for the grism spectroscopy and 0.9ks for the imaging. Data reduction was conducted using the grizli software \citep{Brammer:23}. A full description of the methods will be presented in Brammer et al. in prep, but we summarize here for completeness. The rate files from MAST are aligned to a Gaia-matched reference frame; the direct images are then used to align their associated grism exposures. Bad pixels are masked, then individual exposures are combined. 

We use a running median filter to subtract all continuum emission, leaving a frame with just emission lines. This method subtracts both the continuum from nearby sources that contaminates the spectrum as well as the continuum emission from the source of interest \citep{Kashino:22,matthee:23}. This empirical method has the significant advantage of simplicity. However, it cannot be used if the continuum is of scientific interest, may over-subtract nearby emission lines (e.g. [NII]), and does not model underlying Balmer absorption. The continuum-filtering is a two step procedure. For the first pass, the continuum is subtracted from each row of the image using a running median filter with a 12-pixel central gap to avoid self-subtraction of emission lines. On the second pass, pixels with significant line flux are identified and masked then the median filtering is run again. The position of the spectrum on the detector is computed based on the F444W image and spectral trace in the v4 grism configuration files. 

In addition, imaging in 8 filters (F090W, F115W, F150W, F200W, F277W, F356W, F410M, F444W) was conducted as part of the JWST Advanced Deep Extragalactic Survey (JADES; \citealt{eisenstein:23}). Data are reduced as described in \citet{Tacchella:23} and \citet{rieke:23}. Specifically, the standard \texttt{jwst} pipeline was used with the CRDS context map \texttt{jwst\_1039.pmap}. We used customized procedures to correct for ``wisps'' and 1/f noise, and a background subtraction has been performed both on the individual exposure level and the full mosaics. The mosaic images have been aligned to Gaia-EDR3 catalog. 

Photometry is performed on HST and JWST images Point Spread Function (PSF)-matched to the NIRCam F444W filter. Multi-wavelength catalogs are generated using SExtractor \citep{bertin:96}. Sources are detected in dual-image mode using F444W as the detection image. Fluxes are measured in 0.16" aperture radius and corrected to total using the AUTO flux measurement and a small additional correction for the remaining flux outside this aperture based on the encircled energy for the PSF. All fluxes are corrected for Milky Way foreground extinction using \cite{fitzpatrick:07}. 
Uncertainties on the fluxes are measured from the rms map and multiplied by a scale factor to account for noise not accounted for in the reduction pipeline.  In order to determine this scale factor, we place circular apertures in empty regions of the image and computing the scatter among the flux measurements \citep[see e.g.][]{whitaker:11,skelton:14}. 

Stellar population properties are derived using \texttt{prospector} \citep{leja:17, leja:19,johnson:21} with redshifts set to those inferred from the location of the \ha\ emission line in the grism spectrum of each galaxy. We adopt a 19 parameter physical model that includes redshift, stellar mass, stellar and gas-phase metallicities, star formation history, a two component dust model \citep{charlot:00}, and emission from active galactic nuclei \citep{naidu:22}. We use FSPS \citep{conroy:09} with MIST stellar models \citep{Choi:16}. Star formation rates are computed from the measured \ha fluxes assuming a \cite{chabrier:03} initial mass function with \cite{kennicutt:98}. The fluxes are corrected for dust attenuation using empirical relations based on the UV slope \citep{shivaei:20}. 

Fig.\ref{fig:sfms} shows the distribution of all galaxies with clearly detected \ha\ in the SFR-\mstar\ plane. A locus of points is clearly detected implying the existence of the star-forming main sequence at $z\sim5$ as shown in \citet{shapley:23}. 
The galaxy which is the focus of the present study has log(\mstar/\msun)=10.4 and SFR=185\msun/yr placing it at the high mass end of this distribution and on the star-forming main sequence at this epoch.

We use the \texttt{GALFIT} software package \citep{peng:02, peng:10} to fit the size and shape of this galaxy  accounting for the PSF following the procedure described in \citet{suess22c} and \citet{nelson:23}.  This fitting is performed on the direct images and the \ha\ map (see \S\ref{kinematics}). An empirical PSF is constructed by stacking images of isolated point sources using EPSFBuilder in Photutils \citep[see][for more details]{2023arXiv230518518J}. We create a segmentation map to identify all sources to be modeled or masked. Galaxies that have centers within 3\arcsec  of the target galaxy center and are less than 2.5~mag fainter are modeled simultaneously. Fainter and more distant galaxies are masked. With all sufficiently bright galaxies identified, we estimate and subtract the background in each stamp using the SExtractor background algorithm as implemented in \texttt{photutils}. For F444W, we find effective radius $r_e=2.2$kpc, sersic index $n=1.75$, and axis ratio $q=0.5$. For the medium-band-derived \ha map, we find $r_e=2.3$kpc, $n=0.2$, and $q=0.25$.

\begin{figure*}
    \centering
    \includegraphics[width=0.75\textwidth]{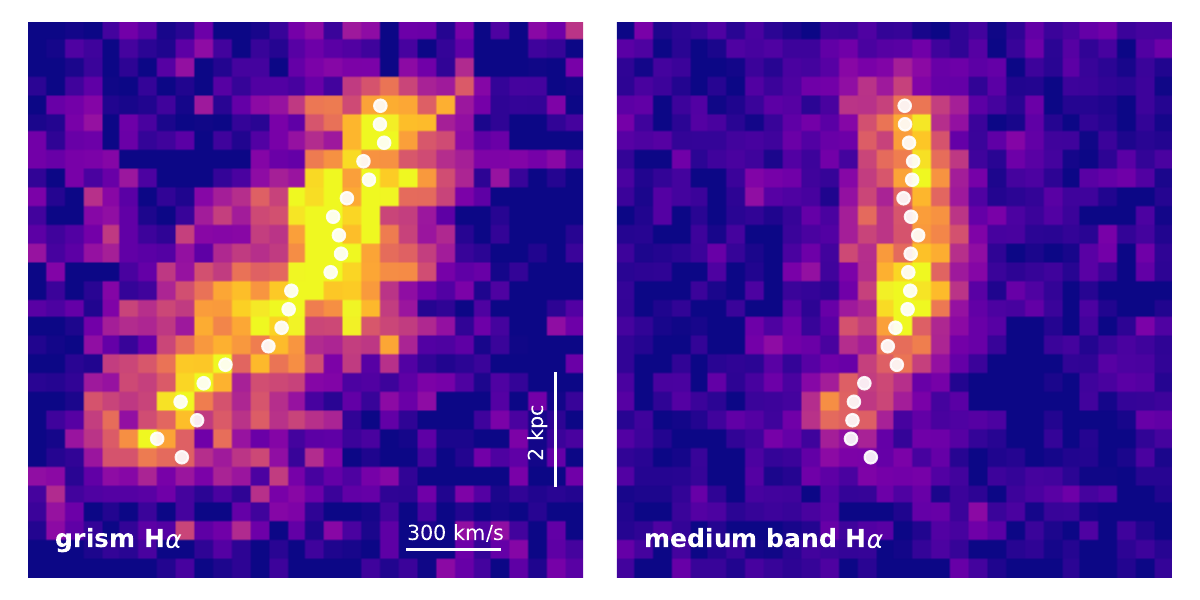}
    \includegraphics[width=0.8\textwidth]{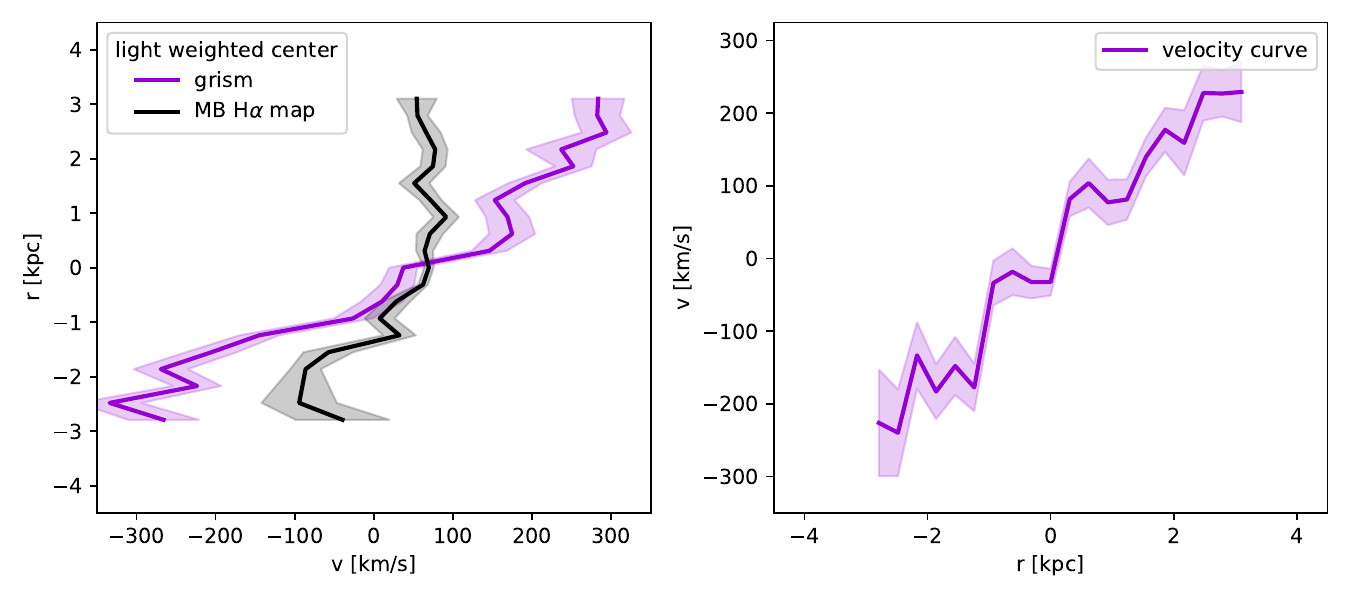}
    \caption{Velocity gradient.  As  described  in \S \ref{kinematics}, we fit the light weighted center along the x-axis in the grism image, which corresponds to the dispersion direction (left), F444W direct image (center), and a stack of the rest-frame UV emission (right).  
    The grism has a spectral resolution of $R\sim1600$, meaning that the grism spectrum contains both spatial and spectral information across in the dispersion direction. The difference between the grism and direct image centroids is the velocity in that spatial pixel. The error bars on the velocity curve are given by the difference between using the rest-UV emission and F444W direct image.}
    \label{fig:kinmodel}
\end{figure*}

\begin{figure*}
    \centering
    \includegraphics[width=0.75\textwidth]{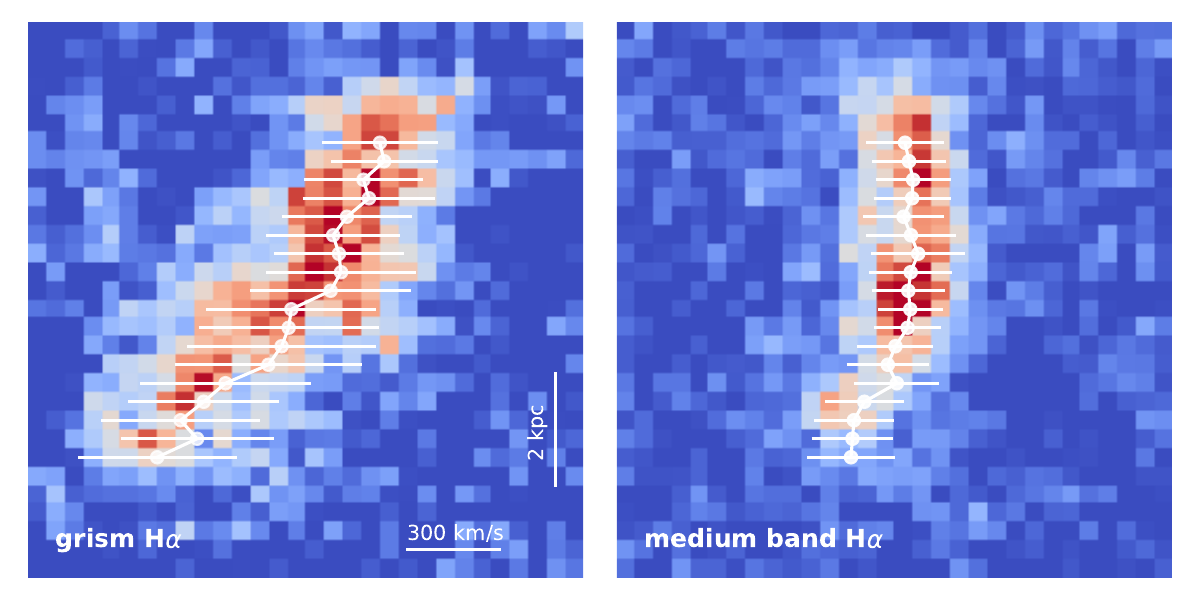}
    \includegraphics[width=0.8\textwidth, trim={0.5cm 9cm 0 0},clip]{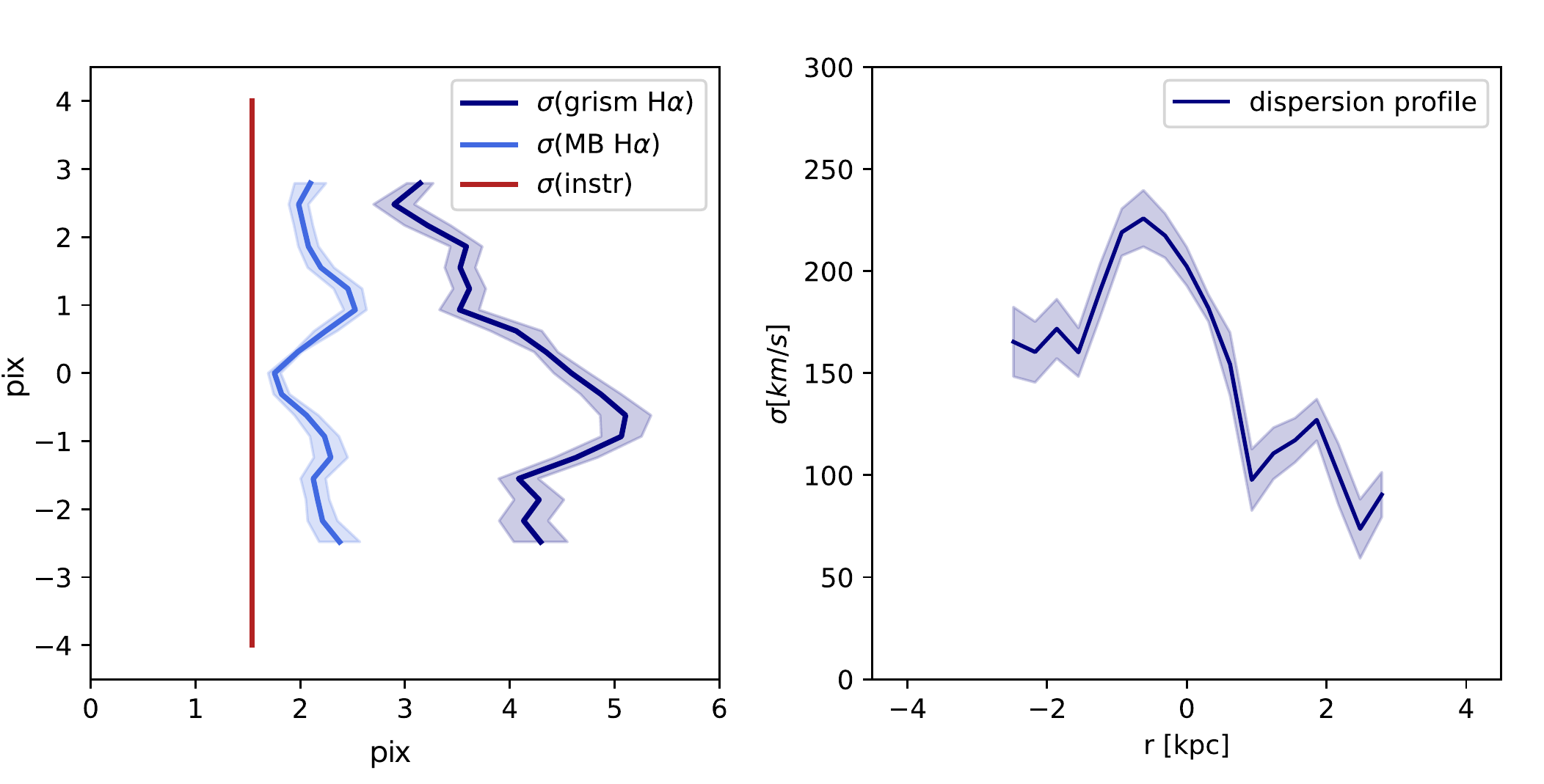}
    \caption{Velocity dispersion profile. As described in \S\ref{kinematics}, the width of the \ha\ line in the spectral direction is broadened by three things: the line-of-sight velocity dispersion of the \ha\ gas (including beam smearing), the spectrograph line spread function, and the intrinsic spatial extent. To measure the line-of-sight velocity dispersion, we subtract the latter two terms off in quadrature. \twister has a centrally peaked velocity dispersion co-spatial with centrally concentrated light, perhaps indicative of a bulge. We measure a disk velocity dispersion of 50-150~km/s, which is likely an upper limit because dynamical forward-modeling has not been performed. The fact that the velocity dispersion peaks in the center and is lower in the disk suggests that the observed velocity gradient is likely due to rotation as opposed to a merger (for which one would expect peaks in the velocity dispersion on either side.}
    \label{fig:sig}
\end{figure*}

\section{Kinematics}\label{kinematics}

A benefit and drawback to slitless grism spectroscopy, depending on the science goal, is that the dispersion axis contains not only spectral information, which is typical, but also spatial information. For HST/WFC3 or JWST/NIRISS the spectral resolution of $R\sim100$ corresponds to  $\sim$1000km/s. In the low resolution regime, both axes of the emission line distribution are effectively spatial and the continuum subtracted line emission is essentially just an emission line map \citep[e.g.][]{vanDokkum:11,Nelson:12,Nelson:13,Nelson:16,Brammer:12,Matharu:21,Matharu:22,Matharu:23}.
The exception is broad-line active galactic nuclei (AGN) with line widths of $> 1000$km/s which can be identified by their compact distribution in the spatial direction and extended distribution in the spectral direction \citep{Nelson:12}.
However, the spectral resolution of the NIRCam F444W grism is much higher -- $R\sim 1600$ at $4\mu$m -- corresponding to a velocity dispersion of $\sigma\sim 80$ km/s. This resolution means that for galaxies with velocity gradients of $>80$km/s, we can place constraints on their kinematic properties in addition to the spatial distribution of their line emission. 

Because the morphology of the emission line along the grism spectral axis is due to both the spatial and kinematic properties, additional data are needed to break the degeneracy. Here we use the F410M medium band which covers the short wavelength portion of the F444W filter in which  the \ha\ line of this object falls. As shown in Fig.\ref{fig:mbha}, the difference between the F410M and F444W image essentially provides a map of the \ha\ line emission \citep[e.g. ][]{williams:23,withers:23}. 
To model the kinematics, we fit the light-weighted center of the \ha\ emission in each row of both the medium band-derived map and the 2D grism spectrum. These datasets share a spatial centroid so the measured difference in light-weighted center is due to velocity. Hence our velocity gradient is inferred by subtracting the medium band \ha\ map light-weighted center from the grism image light-weighted center in each row. 
This method is shown pictorially in Fig. \ref{fig:kinmodel} alongside our results. The uncertainty on the velocity in each spatial pixel is calculated by bootstrap resampling the grism and medium band images within the noise.  

We measure an observed velocity difference $v_{obs}$ from the maximum and minimum velocities by 
$$v_{obs} = \frac{1}{2} (v_{max} -v_{min}) = 235 \pm50 \textrm{ km/s}$$ 
as in \cite[e.g.][]{wisnioski:15}.
If this velocity gradient represents rotation (more on this below), the observed velocity difference must be corrected for inclination in order to reflect the intrinsic rotation velocity:
$$v_{rot} = v_{obs} / \textrm{sin} i$$
We use the axis ratio measured from the medium band \ha\ image (see \S \ref{sec:data}) to infer the inclination of the galaxy using $$\textrm{cos}^2i = (q^2 - q_0^2)/(1-q_0^2)$$ 
where $q=0.25$ is the measured axis ratio and $q_0$ is the intrinsic axis ratio. As we do not in fact know the intrinsic axis ratio, we include the full range of possible intrinsic axis ratios in our error budget $0.05<q_0<q$. Because the measured axis ratio is 0.25, this correction is very small in practice  regardless of the assumed $q_0$ ($<$5\%). 
We find $v_{rot} = 242 \pm50$~km/s.

With the combination of a medium band \ha\ map and grism spectrum, it is also possible to measure the line-of-sight velocity dispersion. The width of the line in the spectral direction is broadened by three things: the spectrograph line spread function, the line-of-sight velocity dispersion of the \ha\ gas (including beam smearing), and the intrinsic spatial extent. 
The contribution of the line spread function is well-known and the degeneracy between the velocity dispersion and spatial extent can be broken using the combination of the medium-band and grism data. 

We fit a Gaussian to the emission line in each row of the \ha\ in both the medium-band map and grism image. These widths are shown in Fig. \ref{fig:sig}. To infer the line-of-sight velocity dispersion, we subtract off the instrumental line spread function and spatial width in quadrature. As shown in Fig. 4, the galaxy has an average velocity dispersion of $\sim100$km/s in the disk and $\sim225$km/s in the center. The most obvious physical explanation for the higher central velocity dispersion is a dynamically hot bulge. Beam-smeared rotation could also play some role, but we note that the measured rotation curve is close to linear. The ratio of rotation to velocity dispersion in the disk, an oft-used metric for the dynamical state of a system is $V/\sigma\sim2$. 

As discussed in \cite{simons:17}, an observed velocity gradient can be an indication of either a rotating disk or a merger. If the object is a disk, one expects centrally peaked velocity dispersions while in a merger the velocity dispersion is expected to be higher on either side and dip in the center. Because we see a centrally peaked velocity dispersion profile and no strong morphological indications for a merger, we interpret the observed velocity gradient as rotation but note that \twister\ may have experienced a merger in the past and whether it is virialized is uncertain.

We infer total dynamical mass of the system using our measured kinematics following \citet{price:22}. 
We compute the circular velocity $V_{circ}$ accounting for turbulent pressure support with 
$$V_{circ} =  (V_{rot}^2 + \alpha\sigma_0^2)^{0.5}$$
\citep[e.g.][]{burkert:10,wuyts:16,wisnioski:18,forster-schreiber:18} where $\alpha = 3.36R/R_e$. 
The dynamical mass within the effective radius of this galaxy, if it can be described by a rotating disk, is  $$M_{dyn}=k(R)\frac{V_{circ}^2(R)R}{G}$$
The virial constant $k(R)$ is dependent on the distribution of mass within the galaxy; we adopt $k(R=R_e) = 2.128$ which is the virial coefficient evaluated for $q=0.4$ and $n=1$ which invokes an elevated $q$ to account for the spherical halo \citep[as in e.g.][]{miller:11,price:20}.
We include the range of velocity dispersions measured in the disk in our dynamical mass uncertainty budget. 

\begin{figure*}
    \centering
    \includegraphics[width=\textwidth]{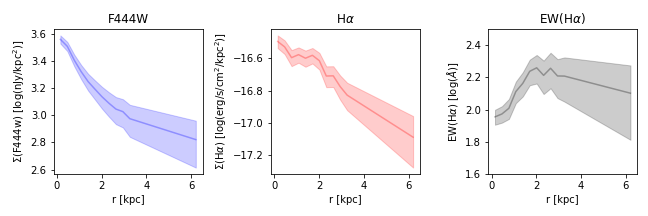}
    \caption{Radial surface brightness profiles of F444W (left), \ha (center), and  the radial \ha\ equivalent width profile (right). The \ha\ distribution is much less centrally concentrated than the 4.4\micron\ continuum emission resulting in an \ha\ equivalent width profile that rises from the center to 2kpc consistent with this $z=5.3$ galaxy growing inside-out in a bulge-disk system.}
    \label{fig:profs}
\end{figure*}

We find a dynamical mass of ${\rm log(M}_{dyn})=11.0\pm0.2$. With a stellar mass of log(M$_*$)=$10.4\pm0.4$, \twister\ has a significantly larger dynamical mass than stellar mass within the effective radius. This stands to reason as the dynamical mass includes the dark matter and gas masses in addition to the stellar mass. We infer a gas mass of ${\rm log(M}_{gas}) = 10.4$ by scaling the SFR with the Kennicutt-Schmidt relation using the effective radius of the \ha emission \citep{kennicutt:98}. Gas fractions are expected to rise dramatically to high redshift \citep{Tacconi20}, so a gas fraction of 50\% is not surprising. Using these estimates of the mass components we infer a baryonic mass that is $\sim50$\% of the total dynamical mass. At $1<z<3$ many studies of ionized gas or stellar kinematics find that massive galaxies ($\log(M_{\star}/M_\odot)\gtrsim10$) are baryon-dominated within the galaxy scale, sometimes with stellar masses alone that match or even exceed their dynamical masses \citep[e.g.][]{forster-schreiber:09,bezanson:13,vandokkum:15,alcorn:16,burkert:16,stott:16,wuyts:16,lang:17,genzel:17,barro:17,forster-schreiber:18,price:20}. 
On the other hand, the first study of the ionized gas kinematics of $z>5$ galaxies with JWST finds dynamical masses an order of magnitude larger than stellar masses, for low mass galaxies ($\log(M_{\star}/M_\odot)\sim7-9$; \citealp{degraaff:23}). 
Larger samples of $z>3$ galaxies covering a range of stellar masses and sizes will be needed to systematically study the baryon-to-dark matter ratios of early disks.

\begin{figure*}
    \centering
    \includegraphics[width=\textwidth]{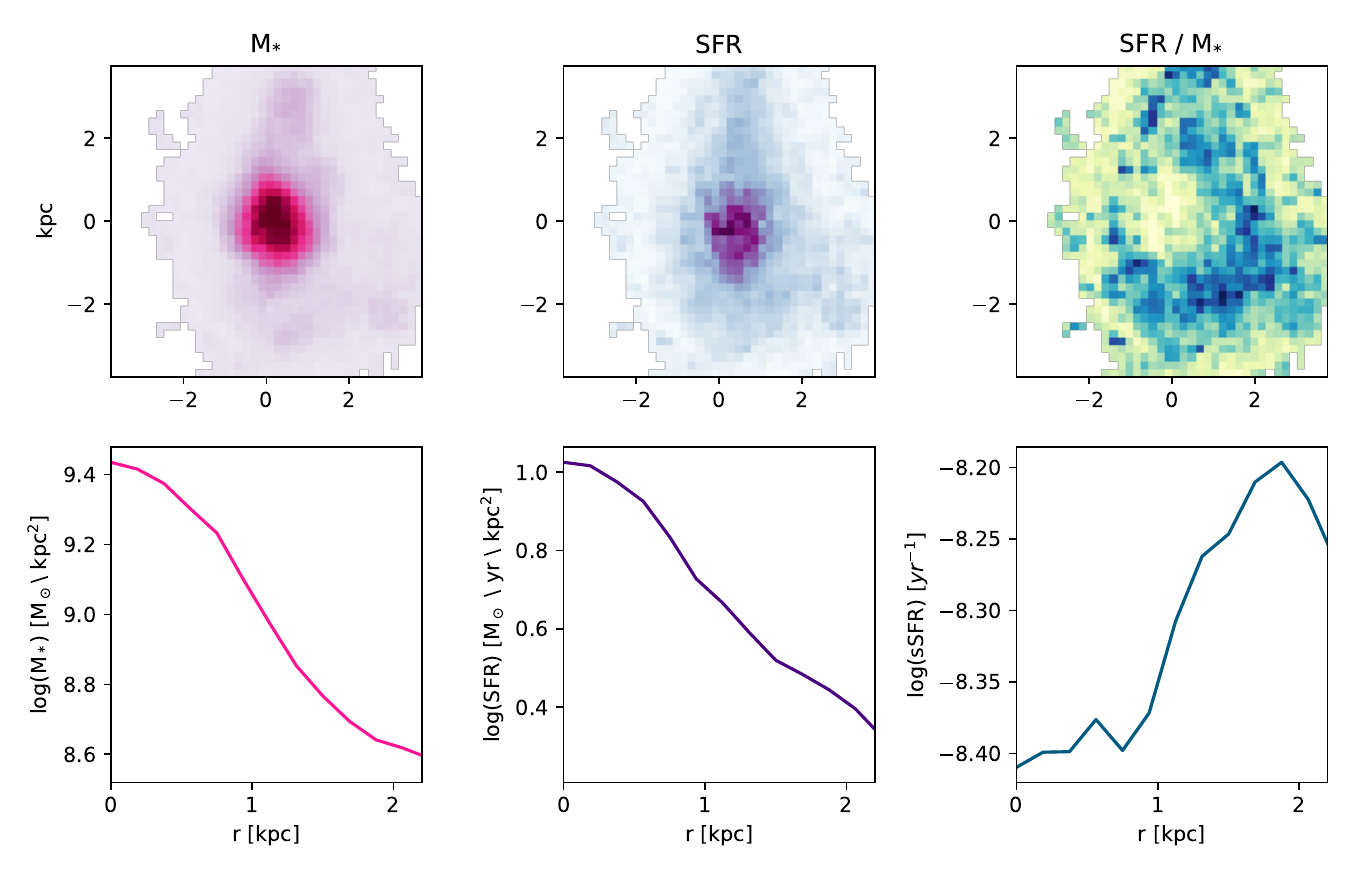}
    \caption{Resolved stellar populations based on SED fitting of the JWST imaging. The top row shows maps of the stellar mass (left), star formation rate (middle), and specific star formation rate (right) from spatially resolved SED modelling of the JWST images. The bottom row shows the radial surface density profiles of the stellar mass, star formation rate, and specific star formation rate, respectively. A clear central depression can be seen in the specific star formation rate, a feature that is typically seen in bulge-disk systems growing inside-out. }
    \label{fig:sps-res}
\end{figure*}

\section{The spatial distribution of H$\alpha$: Inside-out growth}
At $z\sim5-6$, the F444W grism captures \ha\ emission, a tracer of star formation, and the F444W direct image captures the rest-frame R-band continuum, a reasonable proxy for stellar mass. As such, their quotient, the \ha\ equivalent width (EW(\ha)) can serve as a proxy for the specific star formation rate (SFR/M$_*$) which shows how the galaxy is growing \citep[e.g.][]{Nelson:13}. 

Owing to the impact of velocity on the observed morphology of \ha, we cannot measure a radial surface brightness profile in elliptical apertures. Instead, we effectively collapse the light distribution in both the direct and grism images along the detector axis corresponding to the dispersion axis in the grism image. In practice, to optimize the signal-to-noise ratio of our profiles, we instead shift each row to the light weighted centroid and sum along each row encompassing the effective semi-minor axis. The resulting profiles are shown in Fig.\ref{fig:profs}. 

The F444W light is much more centrally concentrated than the \ha\ emission and hence the \ha\ equivalent width profile rises sharply from the center to the effective radius at 2.25~kpc. Taking the EW(\ha) as a proxy for the specific star formation rate, the rising equivalent width we observe suggests that this galaxy is building mass more rapidly at $r\sim r_e$ than in the center. Further, the F444W profile is significantly steeper than the \ha\ profile; PSF effects would only accentuate this difference as the PSF is slightly broader in F444W than F410M. Because the \ha and continuum have the same wavelength, the attenuation they experience from diffuse dust is expected to be the same. While extra attenuation is expected toward HII regions \citep[e.g.][]{price:14,reddy:15}, a variation in the quantity of extra attenuation toward HII regions would need to change as a functon of radius to create the rising EW(\ha) observed. Given that this extra attenuation is surrounding birth clouds, this kind of geometric effect seems unlikely but resolved measurements of H$\beta$ would be needed to be sure \citep[e.g.][]{Nelson:16b,Lorenz:23,shapley:23}. Further, even in the presence of significant dust attenuation, the \ha\ equivalent width is a reasonably good proxy for the sSFR \citep[e.g.][]{nelson:19}. The most obvious interpretation of these features is that this galaxy is growing inside-out with a star-forming disk surrounding a bulge with relatively suppressed star formation. The surprise of course is that this is not $z\sim1$ where this structured inside-out growth is typical \citep[e.g.][]{Nelson:16} but $z>5$.

 We also perform spatially resolved SED fitting of the JWST imaging to determine if the conclusions hold with more sophisticated modeling following \citet{gimenez-arteaga:23}. Briefly, images in all JWST filters are PSF-matched to the F444W resolution and the SED of each pixel is fit. Fits are performed with \texttt{BAGPIPES} \citep{carnall:18} with SPS models by \citet{bruzual:03}, nebular emission with CLOUDY \citep{ferland:17}, a \citet{kroupa:01} initial mass function (IMF), a \citet{calzetti:00} attenuation curve, and constant star formation history following \citet{carnall:23} and \citet{gimenez-arteaga:23}. The redshift is fixed to the grism spectroscopic redshift. Visual extinction is allowed to vary $A_V=0-3$, the maximum age 1Myr - 5Gyr, and metallicity $Z=0-Z_\odot$ all with uniform priors. The resulting maps of stellar mass, star formation rate, and specific star formation rate are shown in Fig.~\ref{fig:sps-res}. As in the case of \ha\ equivalent width, the inferred specific star formation rate is lower in the center than at larger radii, suggesting that this galaxy is building from the inside-out.


\section{Discussion}
We measure kinematics and spatially resolved \ha\ emission for a ${\rm log(M}_*/M_\odot) = 10.4$ galaxy using FRESCO JWST/NIRCam spectroscopy combined with JADES medium band imaging. We measure the effective radius of the rest-frame optical (0.55$\micron$) emission in the F356W filter, which is not expected to be dominated by line emission.  This galaxy has a rest-optical effective radius of $r_e=2.25$ kpc, comparable to the half-light radii of similar-mass galaxies at $z=0.5-1$ \citep[e.g.][]{vanderwel:23}, much larger than we anticipate for an early galaxy if galaxy sizes scale with the scale factor of the universe. It is also significantly larger than one would expect based on existing measurements of the effective radius of galaxies as a function of redshift ($r_e(z)$), which finds  $r_e\sim1$kpc at $z=5.3$ \citep{holwerda:207}.
It has a rising EW(\ha) profile, a proxy for specific star formation rate, and appears to be a bulge-disk system growing inside-out. We measure a rotation velocity of $240\pm50$km/s, similar to or greater than local massive spiral galaxies. 
In many ways this appears to be a typical $z\sim2$ disk galaxy. 
Oddly, it is not at $z\sim2$, but $z=5.38$. 
That massive, extended, rotating disk galaxies could exist at such early times is surprising. In particular because this does not appear to be a one-off situation but rather adds to the growing body of evidence for massive, extended, rotating disks seen in some studies of [CII] with ALMA \citep[e.g.][]{hodge:12, neeleman:20,Tsukui:21,parlanti:23}. 

Considering the dynamical mass with respect to the stellar mass, \twister\ has a stellar mass fraction of $\sim25$\%, and a baryonic mass fraction of $\sim50$\% within the effective radius. With a dynamical mass of log(M$_{dyn}$/\msun)=11.1, these mass fractions are similar to galaxies with similar dynamical masses at $z=1-3$ \citep[e.g][]{forster-schreiber:18,price:20}. If this is confirmed by larger samples, it may mean that the central baryon fraction of massive galaxies is more-or-less set by $z\sim5$. A number of studies find that galaxies become more baryon-dominated in their centers going from $z\sim0$ to $z\sim3$ \citep[e.g.][]{ubler:18,price:20}.
However, at dynamical masses of $10^{11}$\msun, the ratios of stellar-to-dynamical mass and baryonic-to-dynamical mass remain fairly constant \citep[see e.g. Fig 4.][]{price:20}. On the other hand, low  mass, high redshift galaxies are possibly strongly dark matter dominated \citep{degraaff:23}. More massive galaxies at $z\sim7$ have been suggested to show a range in their baryonic to dynamical mass ratio depending on the luminosity-weighted age \citep{Topping:22}, although these measurements are based on spatially unresolved measurements of the [CII] emission line widths and therefore may be affected by mergers and non-virial motions. One interpretation of this result is that the total mass may be more important than the cosmic epoch for determining how baryon dominated the centers of galaxies are. Some theoretical models find that galaxies transition from prolate to oblate when their centers become baryon dominated \citep[e.g.][]{ceverino:15}, consistent with the rotation-dominated kinematics we find. 


We find $V/\sigma\sim2$ for \twister, meaning that although it is rotation dominated, the velocity dispersion in the disk is much higher than that of local galaxies. This is likely to be an upper limit on the velocity dispersion, however, as we do not forward-model the velocity field \citep[as in e.g.][]{degraaff:23} which can result in velocity dispersions which are biased high \citep[e.g.][]{rizzo:22}. 
That caveat noted, a mix of rotational and dispersion support is qualitatively consistent with many recent studies of \ha kinematics at the peak of the cosmic star formation history at $z=1-3$ which have found higher high velocity dispersions in the disks of higher redshift galaxies \citep[e.g.][]{wisnioski:15,swinbank:17,turner:17,ubler:19}. A number of these studies find that $V/\sigma$ declines with redshift to $\sim1$ at $z\sim3$. Based on analytical modeling, \cite{ubler:19} suggest that the significant levels of turbulence in disks at $z>1$ is driven by a combination of gravitational instabilities and stellar feedback \citep[e.g.][]{krumholz:18,varidel:20}. 
Quantitatively, the rotation dominance we observe is higher than one would expect based on the extension of lower redshift \ha studies. \cite{stott:16} and \cite{turner:17} find that the rotation dominated fraction, which they define as $V/\sigma>1$, declines with redshift as $RDF \propto z^{-0.2}$. If extrapolated, this relation implies the likelihood of finding a rotation dominated galaxy at $z\sim5.3$ is less than 10\%. $V/\sigma\sim2$ is, however, consistent with an extrapolation of the $V/\sigma$ values predicted in the TNG50 cosmological hydrodynamical simulation \citep{pillepich:19}. On the other hand, it is significantly lower than the values found by some studies of $z\sim4-5$ galaxies in [CII] emission with ALMA which find $V/\sigma\sim10$ in some massive galaxies at this epoch \citep[e.g.][]{rizzo:20,rizzo:21,fraternali:21,lelli:21}. It is possible that the discrepancy between ionized and molecular gas kinematics is due to actually different kinematics in different gas phases \citep[see e.g.][]{ubler:19}. However, in a detailed study of \ha vs. CO kinematics in a $z=1.4$ galaxy, \cite{ubler:18} find the two tracers yield comparable velocity dispersions. With the objects having kinematic measurements in any tracer at $z>5$ numbering less than 50, resolving this discrepancy will require kinematics for much larger samples in multiple tracers. 

Although the kinematics of \twister\, appear typical for an extrapolation of cosmological simulations, the structure does not. The axis ratio distribution in the TNG50 simulation skews toward higher values than observed galaxies suggesting it has fewer prominent disks than the real universe \citep{kartaltepe:23}. Interestingly, there are almost no objects  at $5<z<6$ in TNG50 with $n>2$ suggesting that bulge-disk systems such as this are not expected in large numbers. Similar mass galaxies in the TNG50 cosmological hydrodynamical simulation have a median observable effective radius of $\leq 1$ kpc \citep{costantin:23}, making an observed effective radius of $r_e=2.25$kpc much larger than expected. There are very few (if any) galaxies with $r_e>2$, $n=2$, and $b/a=0.4$ -- i.e., a spatially extended bulge-disk system such as this. A number of studies find or predict that galaxies form bulges at early cosmic times then build an extended stellar structure around it later \citep[e.g.][]{bezanson:09,nelson:14,zolotov:15,vandokkum:15,wellons:15}, which is qualitatively consistent with what we see in the the structure, kinematics, and particularly the sSFR gradient of \twister. However, a number of studies suggest that building an extended disk is hard at \textit{such} early times. Disks at early cosmic epochs are expected to grow slowly, with formation times of $\sim1.5$Gyr, due to instabilities which drive gas and stars to the center \citep[e.g.][]{ceverino:15,park:19,costantin:22}. Given the mass of this galaxy, it will likely be massive and quiescent by $z\sim2$, placing it in a population of galaxies remarkable for their compactness ($M_*\sim10^{11}M_\odot$ with $r_e\sim1kpc$), suggesting that the disk of this galaxy will likely need to be destroyed.

This study and many others are pointing to the existence of mature galaxies at surprisingly early cosmic epochs. Remarkably luminous galaxy candidates have been found to $z\sim13$ \citep[e.g.][]{Oesch:16,naidu:22,castellano:22,finkelstein:22,bunker:23a,tacchella:23a,harikane:23,donnan:23,casey:23} and candidate massive galaxies with $M_*>10^{10}M_\odot$ have been found at $z=5-9$ \citep[e.g.][]{labbe:23,xiao:23,akins:23}. All of these unexpected sources require spectroscopic confirmation of their nature. Spectra obtained thus far reveal a mixed bag with some redshifts holding \citep[e.g.][]{curtis-lake:23,roberts-borsani:23}, some falling \citep[e.g.][]{arrabal-haro}, and some have remarkable mass or luminosity due to active galactic nuclei instead of stars \citep[e.g][]{Kocevski:23,maiolino:23,matthee:23b}. 
If a number of these results hold, the combination of JWST and ALMA might show us that mature galaxies form earlier than we previously thought possible. This study provides both redshift and dynamical confirmation of a massive galaxy at early cosmic time. 

This paper presents a new method for kinematic measurements with JWST which has the potential to enable dramatic multiplexing for unbiased samples: slitless spectroscopy with medium band imaging. With the fairly high spectral resolution of the NIRCam grism ($R\sim1600$), this mode of spectroscopy could provide an efficient way to derive kinematic constraints on large samples of massive galaxies. Even stronger kinematic constraints could be placed using multiple grism dispersion directions.
While validation tests need to be done against IFU data, this methodology could provide an exciting way to make kinematic measurements for large, unbiased samples of galaxies at early times, allowing us to understand their seemingly rapid early assembly. 


\begin{acknowledgments}
Support for this work was provided by NASA through grant JWST-GO-01895 awarded by the Space Telescope Science Institute, which is operated by the Association of Universities for Research in Astronomy, Inc., under NASA contract NAS 5-26555. H{\"U} gratefully acknowledges support by the Isaac Newton Trust and by the Kavli Foundation through a Newton-Kavli Junior Fellowship. 
This work has received funding from the Swiss State Secretariat for Education, Research and Innovation (SERI) under contract number MB22.00072, as well as from the Swiss National Science Foundation (SNSF) through project grant 200020\_207349. The Cosmic Dawn Center (DAWN) is funded by the Danish National Research Foundation under grant No. 140. RS acknowledges an STFC Ernest Rutherford Fellowship (ST/S004831/1). RPN acknowledges support for this work provided by NASA through the NASA Hubble Fellowship grant HST-HF2-51515.001-A awarded by the Space Telescope Science Institute, which is operated by the Association of Universities for Research in Astronomy, Incorporated, under NASA contract NAS5-26555. MVM acknowledges support from the National Science Foundation via AAG grant 2205519 and the Wisconsin Alumni Research Foundation via grant MSN251397.  RM also acknowledges funding from a research professorship from the Royal Society. AJB, AJC, \& GCJ acknowledge funding from the "FirstGalaxies" Advanced Grant from the European Research Council (ERC) under the European Union’s Horizon 2020 research and innovation programme (Grant agreement No. 789056). IL acknowledges support by the Australian Research Council through Future Fellowship FT220100798. DJE is supported as a Simons Investigator and by JWST/NIRCam contract to the University of Arizona, NAS5-02015. RM, JW, LS, and WB acknowledge support by the Science and Technology Facilities Council (STFC), by the ERC through Advanced Grant 695671 “QUENCH”, and by the UKRI Frontier Research grant RISEandFALL. BER acknowledges support from the NIRCam Science Team contract to the University of Arizona, NAS5-02015. The research of CCW is supported by NOIRLab, which is managed by the Association of Universities for Research in Astronomy (AURA) under a cooperative agreement with the National Science Foundation.  
\end{acknowledgments}

\vspace{5mm}
\facilities{JWST, HST}


\software{Grizli}


\bibliography{ms}{}
\bibliographystyle{aasjournal}



\end{document}